\documentclass[aps,amsmath,amssymb,nofootinbib,superscriptaddress,showpacs,floatfix,prl,twocolumn,balancelastpage,flushbottom]{revtex4-1}

\usepackage{latexsym}
\usepackage{graphicx}
\usepackage{times,psfrag,subfigure}
\usepackage{amsmath}
\usepackage{dcolumn}
\usepackage{bm}       
\usepackage{color}
\usepackage{latexsym,amsmath,amssymb,bm,euscript}
\usepackage{flushend}
\usepackage{balance}

\newcommand{\beq}{\begin{equation}}
\newcommand{\eeq}{\end{equation}}
\newcommand{\beqarray}{\begin{eqnarray}}
\newcommand{\eeqarray}{\end{eqnarray}}

\newcommand{\eq}[1]{Eq.~(\ref{#1})}  
\newcommand{\fig}[1]{Fig.~\ref{#1}}

\begin{document}

\allowdisplaybreaks

\title{Edge currents as a signature of flat bands in topological
  superconductors}

\date{\today}

\author{Andreas P. Schnyder}
\email{a.schnyder@fkf.mpg.de}
\affiliation{Max-Planck-Institut f\"ur Festk\"orperforschung,
  Hei\ss{}enbergstrasse 1, D-70569 Stuttgart, Germany} 

\author{Carsten Timm}
\email{carsten.timm@tu-dresden.de}
\affiliation{Institute of Theoretical Physics, Technische Universit\"at
  Dresden, D-01062 Dresden, Germany} 

\author{P. M. R. Brydon}
\email{brydon@theory.phy.tu-dresden.de}
\affiliation{Institute of Theoretical Physics, Technische Universit\"at
  Dresden, D-01062 Dresden, Germany}

\begin{abstract}
We study nondegenerate flat bands at the surfaces of
noncentrosymmetric topological superconductors by exact
diagonalization of Bogoliubov-de
Gennes Hamiltonians. We show that these states are strongly spin 
polarized, and acquire a chiral dispersion when placed in contact
with a ferromagnetic insulator. This chiral mode carries a large edge current
which displays a singular dependence on the exchange-field strength. The
contribution of other edge states to the current is comparably weak.
We hence propose that the observation of the edge current can
serve as a test of the presence of nondegenerate flat bands.
\end{abstract}

\date{\today}

\pacs{74.50.+r, 74.20.Rp, 74.25.F-, 03.65.vf}

\maketitle

\textit{Introduction}.---The bulk gap of topological insulators and
superconductors plays an essential role in defining the topological
invariants, and hence for the topological protection of their
surface states~\cite{hasanKane2010,ryuNJP10,qiZhangReview2010,schnyder2008}.
Recently, however, the topological classification of matter has been extended
to \textit{gapless} systems, such as Dirac or Weyl
semimetals~\cite{Wan2011,Turner2013} and nodal 
superconductors~\cite{Tanaka2012,Schnyder2011,Matsuura2012,Zhao2012}. A
bulk-boundary correspondence exists for certain surfaces, yielding
topologically protected dispersionless zero-energy states, 
so-called ``arc lines'' or ``flat bands''. Well-known examples are the
zero-energy edge states of
cuprate superconductors and the A phase of $^3$He.

A promising materials class for topological systems is the noncentrosymmetric
superconductors (NCSs), characterized by strong 
antisymmetric spin-orbit coupling (SOC) and a mixing of spin-singlet and
spin-triplet pairing~\cite{bauerSigristbook}. The superconducting gap in many
of these 
compounds is reported to display line nodes, e.g., in
CePt$_3$Si~\cite{Izawa2005,Bonalde2009}, CeIrSi$_3$~\cite{Mukuda2008},
and  Li$_2$(Pd$_{1-x}$Pt$_x$)$_3$B~\cite{Yuan2006,Nishiyama2007,Eguchi2013}, and
they can 
therefore support topological flat-band surface states. Due to the exotic gap
structure of NCSs, these flat bands are predicted to be \emph{nondegenerate},
i.e., Majorana 
fermions~\cite{Tanaka2010,satoyada2011,Schnyder2011,Brydon2011,Schnyder2012,Dahlhaus2012},
in contrast to the doubly degenerate flat bands found in other
systems. Demonstrating that the surface flat bands of an NCS are
nondegenerate presents a challenge, however:
While typical experimental methods, such as tunneling conductance
spectroscopy, are sensitive to the singular surface density of states
contributed by the flat bands, they cannot probe the degeneracy.

In this Letter, we propose the response of the nondegenerate flat bands to a
proximity-induced exchange field as an unambiguous test of their
existence. We lay the foundation for our approach by demonstrating that the
flat-band states are 
strongly spin polarized, which in itself is an important experimental
signature. The spin polarization originates from both the SOC and the 
spin structure of the superconducting gap. Consistent with the
time-reversal invariance of the pairing state, the spin polarization is odd in
the surface momentum.  
Upon bringing the superconductor into contact with a ferromagnetic insulator,
the flat-band states therefore acquire a \emph{chiral} dispersion due to the
coupling to the exchange field, and hence carry a sizable charge current
along the interface. The current displays a remarkable singular dependence on
the exchange-field strength: Only an infinitesimal exchange field 
is required to generate a large current, and small changes in the external
field can switch the current's sign. In contrast, 
doubly degenerate flat bands, when present,
give a much weaker contribution to the
current. We further show that the interface current for a  
\emph{fully gapped} NCS is  small and not simply related to the
presence of spin-polarized edge states.

\emph{Model Hamiltonian}.---%
Quasiparticle motion in an NCS is described
by the Bogoliubov-de Gennes Hamiltonian $H = \sum_{\bf k}\Psi_{\bf
    k}^{\dagger}H_{\bf k}\Psi_{\bf k}$,
 with
$\Psi_{\bf k} = ( c_{{\bf k} \uparrow},  c_{{\bf k} \downarrow},  c^\dag_{-{\bf
k} \uparrow} , c^\dag_{-{\bf k} \downarrow}  )^T$ and
\begin{eqnarray}
\label{TBmodel}
H_{\bf k} = \begin{pmatrix}
\varepsilon_{\bf k} \sigma_0  - \lambda\, {\bf l}_{\bf k} \cdot \bm{\sigma}  &
(\psi_{\bf k} \sigma_0 + {\bf d}_{\bf k} \cdot
\bm{\sigma} ) i \sigma_y \cr
-i\sigma_y(\psi_{\bf k}^\ast \sigma_0 + {\bf d}^\ast_{\bf k} \cdot
\bm{\sigma} )  &  - \varepsilon_{\bf k} \sigma_0
  - \lambda\,  {\bf l}_{\bf k} \cdot \bm{\sigma}^{\ast} 
\end{pmatrix} .\quad
\end{eqnarray}
Here, ${\bm \sigma}$ is the vector of Pauli matrices.
The normal part of the Hamiltonian describes a two-dimensional square lattice
with nearest-neighbor hopping and chemical potential $\mu$, $\varepsilon_{\bf
  k} = 2t\, ( \cos k_x +
\cos k_y) - \mu$, and Rashba SOC with
${\bf l}_{\bf k} = \hat{\bf x}\, \sin k_y 
- \hat{\bf y}\, \sin k_x$ and strength $\lambda$.
The even-parity spin-singlet and odd-parity spin-triplet superconducting gaps
are written as
$\psi_{\bf k} = \Delta_{0} f({\bf k})\, q$ and ${\bf d}_{\bf k} = \Delta_0
f({\bf k})\, {\bf l}_{\bf k}\, ( 1 -q )$, respectively, where the parameter
$q$ tunes between purely spin-triplet
($q=0$) and purely spin-singlet ($q=1$) pairing.
The structure factor $f ( {\bf k})$ fixes the
orbital-angular-momentum pairing state. Here we focus on
the nodal ($d_{xy}$+$p$)-wave state~\cite{Tanaka2010} described by
$f({\bf k}) = \sin k_x \sin k_y$. We also present contrasting
results for the fully gapped ($s$+$p$)-wave
state~\cite{Iniotakis2007,NCSspin,satoFujimoto2009} with
$f({\bf k})=1$.
In our calculations we fix $(t, \mu, \lambda, \Delta_0 ) = (2.0,
4.0, -2.0, 0.5)$. Different values of these parameters do not qualitatively
change our results.

%%%%%%%%%%%%%%%%%%%%%%%%%%%%
\begin{figure}[t!]
\centering
\includegraphics[clip,angle=0,width=\columnwidth]{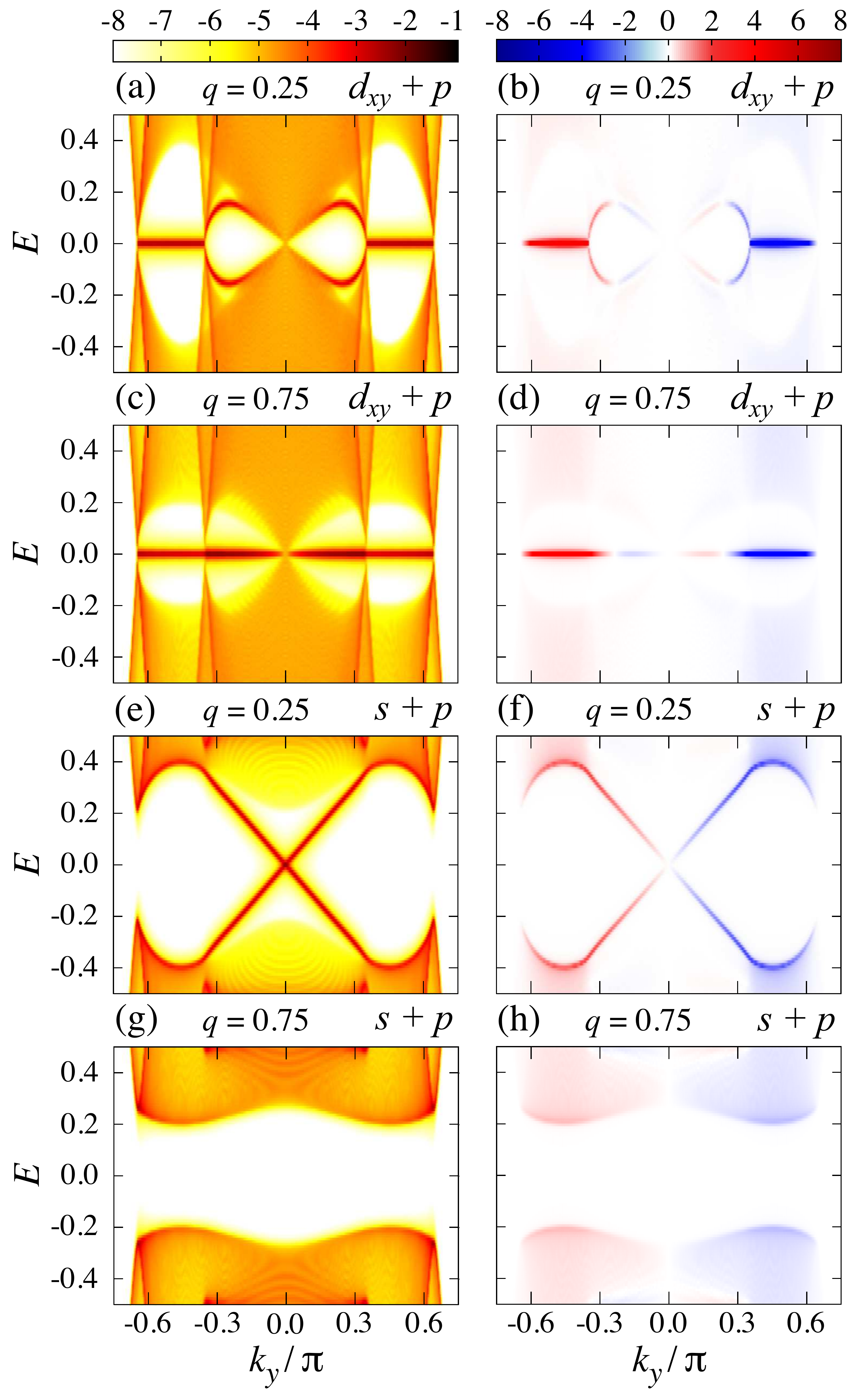}
\caption{\label{fig1L}(Color online)
Left column: Momentum-resolved LDOS on a log scale for the ten outermost
layers at the (10) edge of the ($d_{xy}$+$p$)-wave NCS with (a)
$q=0.25$ and (c) $q=0.75$, and for the ($s$+$p$)-wave NCS with (e)
$q=0.25$ and (g) $q=0.75$. In the right column we present the corresponding
$x$ component of the momentum-resolved spin-LDOS for the outermost layer in
units of $\hbar / 20$ on a linear scale. The $z$ component of the 
momentum-resolved spin-LDOS is provided in Fig.~S1 of the SM~\cite{SM}.}
\end{figure}
%%%%%%%%%%%%%%%%%%%%%%%%%%%%

\emph{Edge states}.---%
The topological properties of the NCS are best revealed by examining
the subgap edge states.
To that end, we compute the spin- and momentum-resolved local density of
states (LDOS) of Hamiltonian~\eqref{TBmodel} in a ribbon of width $L_x$
with edges perpendicular to the (10) direction. 
The momentum-resolved LDOS and spin-LDOS in the
\textit{n}-th layer are given by
\begin{subequations} \label{DOSeq}
\begin{align}
\rho^{\ }_n (E, k_y)
& = - \frac{1}{4 \pi} \, \textrm{Im}
\left[ \textrm{Tr} \left\{
 G_{k_y} ( n , n; E) 
\right\} \right] , \\
\rho^{\mu}_n (E, k_y)
& = - \frac{\hbar}{4 \pi} \, \textrm{Im}
\left[ \textrm{Tr} \left\{
\check{S}^\mu\, G_{k_y} ( n , n; E) 
\right\} \right] ,
\end{align}
\end{subequations}
respectively,
where $\check{S}^\mu = \textrm{diag} \left( \sigma^\mu, {-} [\sigma^\mu]^\ast
\right)$, and $G_{k_y}(n,n';E) =
-i\int^\infty_{-\infty} dt\, e^{iEt}\langle 
T_{t}\,\Psi^{}_{n k_y}(t)\Psi^\dagger_{n'k_y}(0)\rangle$ is the zero-temperature
Green's function with
$\Psi_{n k_y} = (2\pi L_x)^{-1/2}
\sum_{k_x} \Psi_{{\bf k}}\, e^{- i k_x n}$.
The expressions in Eqs.~\eqref{DOSeq} are evaluated for ribbons of width up to
$L_x = 10^3$ and an intrinsic broadening $\eta=0.005$.
Figure \ref{fig1L} shows the LDOS  
integrated over the ten outermost layers, which is on the order of the
localization length of the subgap state. 
Both subgap edge states (dark red/gray) and continuum bulk states
(orange and yellow/light gray) are visible.

The nodal character of the ($d_{xy}$+$p$)-wave state precludes the
existence of a global topological number. By treating every point in the
edge Brillouin zone 
(BZ) as the edge of a one-dimensional system, however, one can define a 
momentum-dependent winding number $W_{\textrm{(10)}}(k_y)$, which only changes
across projected nodes of the bulk 
gap~\cite{Tanaka2012,satoyada2011,Schnyder2011,Schnyder2012}. 
In particular, one finds $W_{\textrm{(10)}}(k_y)=\pm1$ for $k_y$ lying between
the projected edges $k_{F,+}$ and $k_{F,-}$ of the two spin-orbit-split Fermi
surfaces, i.e., $|k_y| \in [k_{F,+}, k_{F,-} ] \simeq [0.352  \pi, 0.648
  \pi]$. This ensures the existence of nondegenerate zero-energy flat
bands at these momenta $k_y$, which are clearly visible in
Figs.~\ref{fig1L}(a) and (c). 
For $|k_y|<k_{F,+}$, on the other hand, there are 
topologically trivial dispersing states for dominant triplet pairing, and
doubly degenerate zero-energy flat bands with $W_{(10)}(k_y)=\pm2$
when singlet pairing dominates. In contrast, for $q<q_{c,L}\simeq 0.472$ and
$q>q_{c,U} \simeq 0.583$, the ($s$+$p$)-wave NCS is a fully gapped
superconductor in symmetry class DIII.
For $q<q_{c,L}$, the superconductor has a topologically nontrivial
character with a non-zero $\mathbb{Z}_{2}$ topological
number~\cite{satoFujimoto2009,ryuNJP10,QiHughesZhangPRB2010}, and by the
bulk-boundary correspondence there are
helical Majorana subgap states, see Fig.~\ref{fig1L}(e) \cite{Iniotakis2007,NCSspin,satoFujimoto2009}. For
$q>q_{c,U}$, the system is topologically trivial and there are no edge
state [Fig.~\ref{fig1L}(g)].

%%%%%%%%%%%%%%%%%%%%%%%%%%%%
\begin{figure}[t!]
\centering
\includegraphics[clip,angle=0,width=\columnwidth]{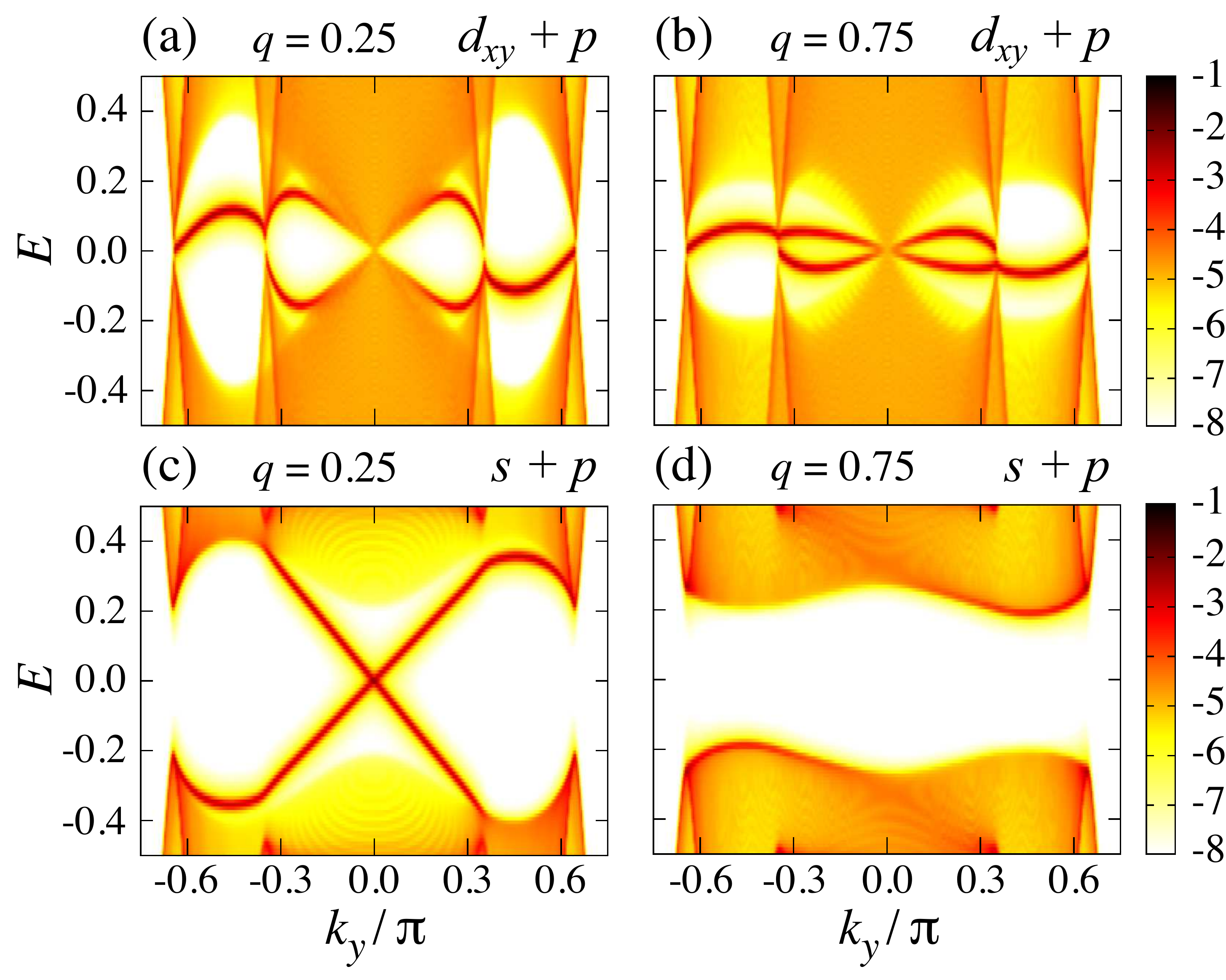}
\caption{\label{fig2L}(Color) 
Momentum-resolved LDOS at the
(10) edge of (a), (b) the ($d_{xy}$+$p$)-wave
and (c), (d) the ($s$+$p$)-wave NCS in the presence of an exchange field
along the $x$ axis with $H^{x}_{\rm ex}=0.4$, applied to the edge layer
$n=1$. In the left column we have $q=0.25$ (majority triplet), while the right
column we plot results for $q=0.75$ (majority singlet). As in 
Fig.~\ref{fig1L} we use a log color scale.}
\end{figure}
%%%%%%%%%%%%%%%%%%%%%%%%%%%%

Similarly to other topological systems with strong SOC
\cite{hasanKane2010,simonPRL12}, the NCS edge states exhibit a distinct spin 
texture. It is well known that the electronlike part of the edge-state
wavefunction  
is strongly spin polarized~\cite{NCSspin}. The \emph{total} spin
  polarization, which includes both electronlike and holelike polarizations,
  is also nonvanishing. Importantly, the exchange field couples to the
    total spin polarization, not just to the electronlike contribution.
In both NCS models we find that the continuum and subgap states show a strong
total polarization in the $xz$ spin-plane, but a vanishing
$y$-component~\cite{yspin}. We present the $x$-spin
polarization in the right-hand column of Fig.~\ref{fig1L}, and the $z$-spin
polarization in Fig.~S1 of the Supplemental Material (SM)~\cite{SM}.
The magnitude and sign of the polarization are
strongly momentum dependent, and display a complicated interplay
between  the singlet-triplet ratio $q$ and the SOC strength $\lambda$.
The nondegenerate zero-energy flat bands of the ($d_{xy}$+$p$)-wave NCS
exhibit a particularly strong and robust $x$-spin polarization,
whereas the $z$-spin polarization is smaller and changes sign close to $\pm
(k_{F,-} - k_{F,+})/2$ \cite{SM}. This is in contrast to the doubly
degenerate states in 
the singlet-dominated state, which give opposite contributions to the
spin-LDOS of unequal magnitude, overall leading to a weaker spin
polarization than for the nondegenerate states.
As required by time-reversal symmetry, subgap states with opposite edge momenta
have opposite spin polarizations. This enhances the robustness of the
nondegenerate flat bands, as time-reversal-invariant scattering processes
connecting the oppositely $x$-spin-polarized states at  $+k_y$ and $-k_y$
are strongly suppressed.

%%%%%%%%%%%%%%%%%%%%%%%%%%%%
\begin{figure}
\includegraphics[clip,angle=0,width=\columnwidth]{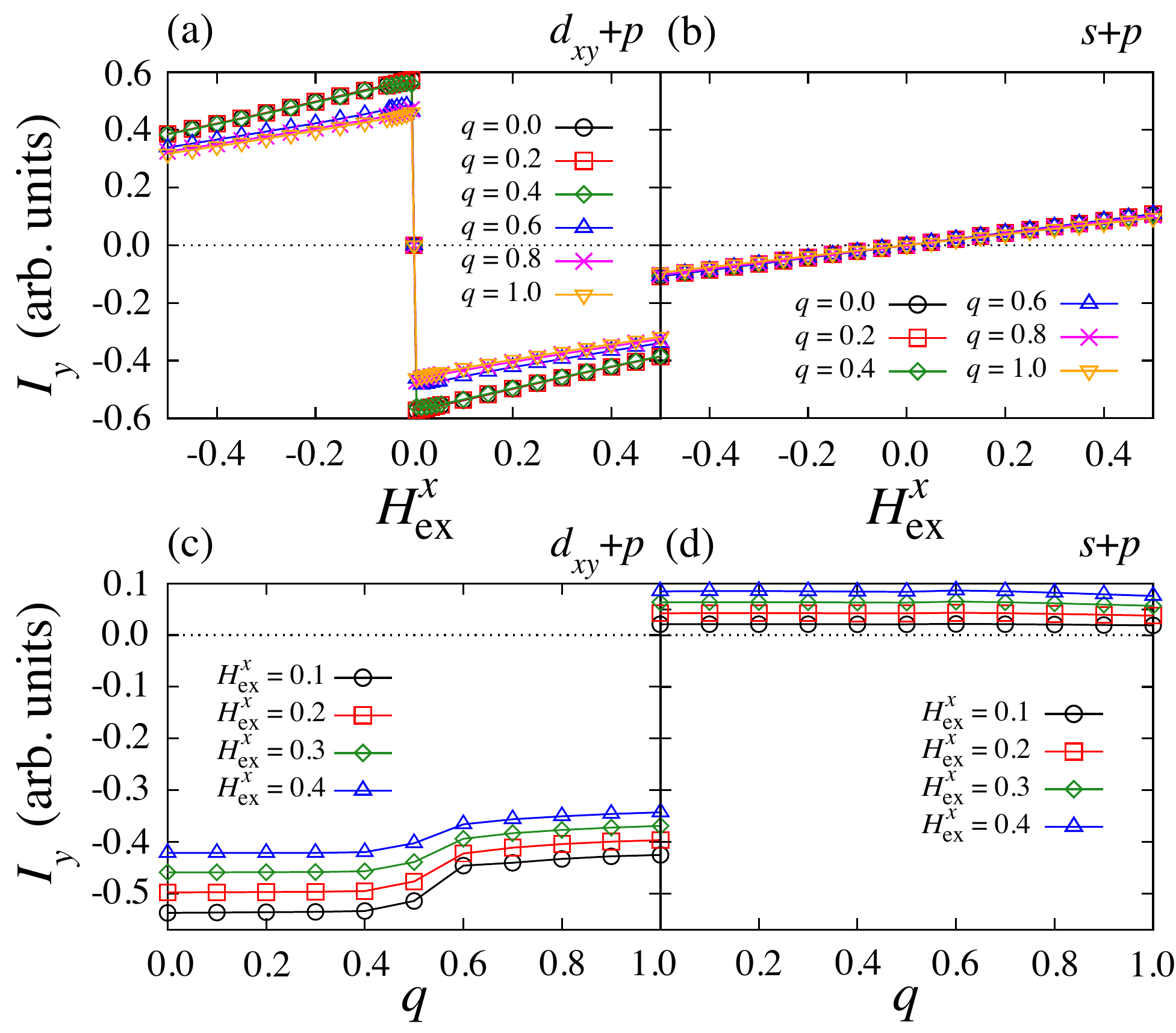}
\caption{\label{fig3L}  (Color online)
Top row: Zero-temperature edge current $I_y$ as a function of the
exchange field $H^{x}_{\rm ex}$ for various values of  the
singlet-triplet parameter $q$ for (a) the ($d_{xy}$+$p$)-wave and 
(b) the ($s$+$p$)-wave NCS. Bottom row: $I_y$ as a function of the
singlet-triplet parameter $q$
for various exchange fields along the $x$ axis,
applied to the leading edge of (c) the ($d_{xy}$+$p$)-wave and (d)
the ($s$+$p$)-wave NCS. For comparison, the edge current of a
chiral $p$-wave superconductor without an exchange field or SOC is about
$0.64$ in our units.}
\end{figure}
%%%%%%%%%%%%%%%%%%%%%%%%%%%%

The nontrivial spin character of the edge states could be inferred
from the absence of large-$k_y$ backscattering processes in quasiparticle
interference measurements. Another possibility is to study the response of
the subgap states upon bringing the NCS into contact with a ferromagnetic
insulator. 
We anticipate that the proximity-induced exchange field ${\bf H}_{\textrm{ex}}$
will lead to a perturbative correction to the energy of the spin-polarized edge
states proportional to  $\sum_{\mu=1}^3 H^{\mu}_{\textrm{ex}}\,
\rho^{\mu}_{1}(E, k_y)$.
Since the flat-band states at $+k_y$ and $-k_y$ are oppositely spin polarized,
the coupling to the exchange field will shift the energy of these edge states
in opposite directions, hence converting them into
\emph{chirally} dispersing modes. Similarly, the left- and right-moving
helical edge states of the ($s$+$p$)-wave NCS should acquire different
velocities. 
To test this, we show in Fig.~\ref{fig2L} the momentum-resolved
LDOS  
when an exchange field
$\mathbf{H}_{\textrm{ex}} = 0.4\,\hat{\bf x}$ is applied to the leading
edge. Here and in the rest of this Letter we will specialize to an
$x$-oriented exchange field;
results for a field along the $z$ axis 
are included in the SM~\cite{SM}.
Comparison with Fig.~\ref{fig1L} reveals that the edge states of
both NCS systems indeed display a linear
shift in energy, which is to a good approximation proportional to 
$H^{x}_{\textrm{ex}}\, \rho^{x}_{1}(E, k_y)$, as long as
$|H^{x}_{\textrm{ex}}| \lesssim \Delta_0$.

%%%%%%%%%%%%%%%%%%%%%%%%%%%%
\begin{figure}[t!]
\includegraphics[clip,angle=-0,width=\columnwidth]{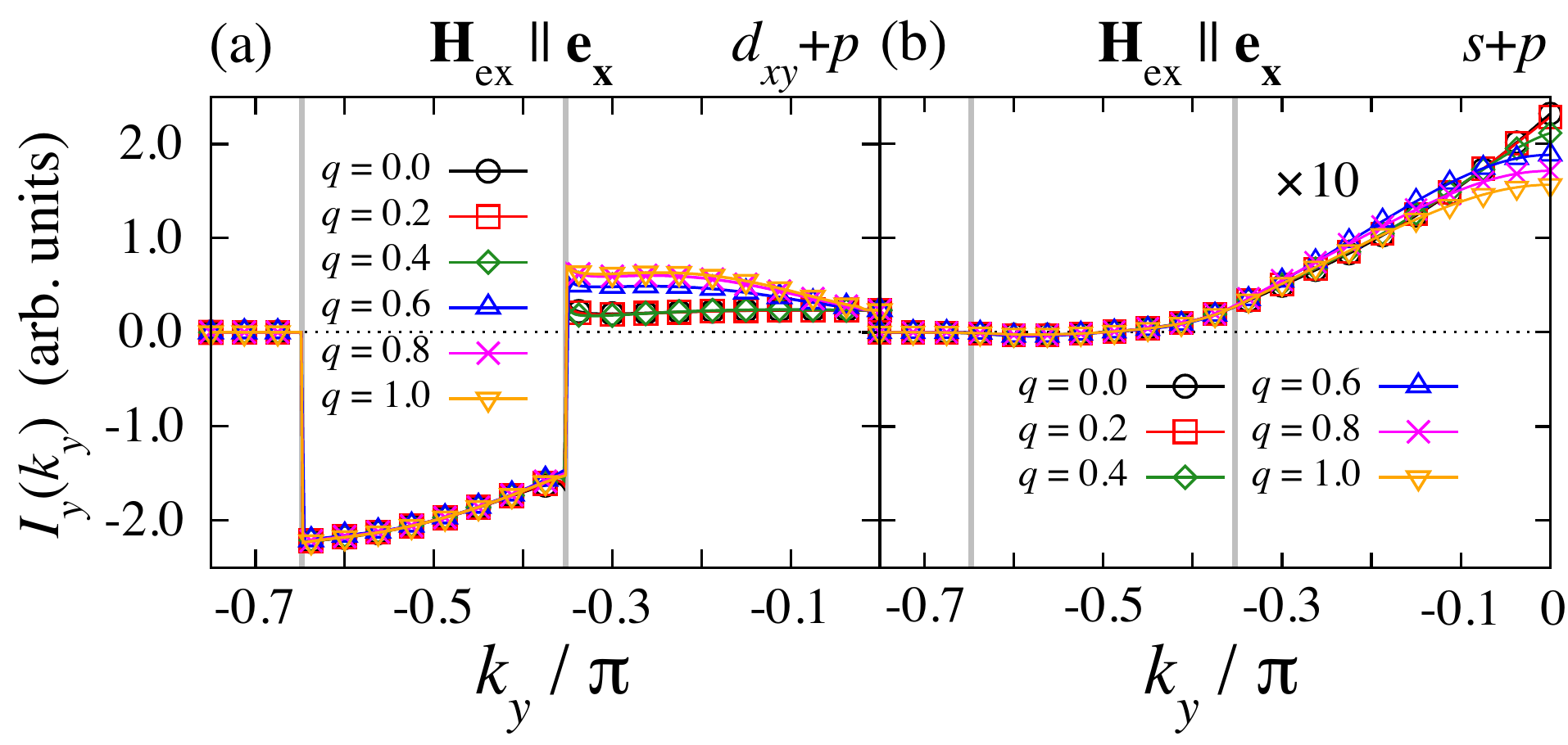}
\caption{\label{fig4L} (Color online)
Momentum-resolved edge current $I_y(k_y)$ for various values of
singlet-triplet parameter $q$ in (a) the ($d_{xy}$+$p$)-wave and 
(b) the
($s$+$p$)-wave NCS. The exchange field is $H^{x}_{\text{ex}}=0.2$. The
vertical gray lines indicate the projections of the edges of the two
spin-orbit-split Fermi surfaces, $|k_y| = k_{F,\pm}$. Note that the
($s$+$p$)-wave results are multiplied by $10$ for clarity.}
\end{figure}
%%%%%%%%%%%%%%%%%%%%%%%%%%%%

\emph{Edge Currents}.---% 
The chiral structure of the edge states induced by the exchange field
  naturally suggests the presence of a spontaneous edge current 
in the NCS. In particular, the chiral mode originating from the flat bands
can be expected to carry a current comparable to that in a chiral $p$-wave
superconductor. 
The zero-temperature expectation value of the surface component
  of the current is written in terms of the momentum-resolved LDOS and
spin-LDOS as
\beqarray
I_y & = &
\frac{e}{2\hbar}\frac{1}{N_y}
\sum_{k_y}\sum_{n=1}^{L_x/2}\int_{-\infty}^{0}dE\, \notag
\\
&& \times \left\{2t\sin k_y \,\rho_{n}(E,k_y) - \lambda\cos
k_y\,\rho^x_{n}(E,k_y)\right\}\,. \label{totcur}
\eeqarray
Here, $N_y$ is the number of $k_y$ points in the edge BZ.
The first term in the braces is the contribution from
nearest-neighbor hopping, whereas the second term is due to the SOC.

In Fig.~\ref{fig3L} we plot the edge current $I^{}_y$ as a
function of singlet-triplet parameter $q$ and exchange field along the
$x$ axis. 
We indeed find spontaneous
currents flowing along the edge of the NCS for both gap symmetries, but  
the two cases are dramatically different.
In particular, the current in the ($d_{xy}$+$p$)-wave NCS exhibits striking
deviations from linear response behavior: As seen in Fig.~\ref{fig3L}(a), an
infinitesimally small exchange field is sufficient to generate a large current
in the NCS, and the current abruptly switches sign as the exchange field is
reversed. Remarkably, the magnitude of the current \emph{decreases} with
increasing exchange field from  a maximum magnitude at $H^{x}_{\rm ex}
\rightarrow \pm0$, which is about $90$\% of the one
for a chiral $p$-wave superconductor with vanishing SOC. 
Although the current is somewhat larger for triplet-dominated
pairing, it depends only weakly on $q$ away from the singlet-triplet crossover
($q_{c,L}<q<q_{c,U}$), as
shown in Fig.\ \ref{fig3L}(c). In contrast, the current in the
($s$+$p$)-wave NCS is always much smaller  and grows linearly with the
exchange field, see Figs.~\ref{fig3L}(b) and (d). Unexpectedly, there is
almost no dependence upon the singlet-triplet parameter $q$, despite the
qualitative change of the subgap spectrum with the topological transition at
$q=q_{c,L}$.

To gain a better understanding of the origin of the currents, it is instructive
to examine the current $I_{y}(k_y)$ contributed by states at
$+k_y$ and $-k_y$, i.e., the even part of the $k_y$ summand
in~\eq{totcur}. In Fig.~\ref{fig4L} 
we show the evolution of $I_y ( k_y)$ with the singlet-triplet
parameter~$q$ for an exchange field along the $x$ axis. The
current in the ($d_{xy}$+$p$)-wave NCS is dominated by the contribution 
from states with $k_{F,+} < |k_y| < k_{F,-}$, i.e., the momenta at
which we find the chiral mode originating from the nondegenerate flat
bands. The current displays no variation with $q$ in this
momentum range and is almost independent of the exchange-field strength (not
shown), and hence accounts for most of the singular response seen in
Fig.~\ref{fig3L}(a). The linear decrease of the total current with increasing
exchange field originates from the contribution at
$|k_y|<k_{F,+}$. Intriguingly, in this region there is also a small singular
response in the singlet-dominated regime, accounting for the reduced jump in
the total current in Fig.\ \ref{fig3L}(a). Its origin will be discussed
briefly below. The profile of
$I_y(k_y)$ in the ($s$+$p$)-wave NCS shows
that the edge current is in this case due to states with $ |k_y|<
k_{F,+}$. Although this is the momentum range in which the helical edge states
are realized for $q<q_{c,L}$, there is little
$q$ dependence of the momentum-resolved currents. 
We hence conclude that the current is largely insensitive to the helical edge
states. 

\emph{Discussion}.---%
The strong edge current in the ($d_{xy}$+$p$)-wave NCS is
primarily carried by the chiral edge mode originating from the
nondegenerate flat bands. Although the exchange field induces chiral
structures in all the edge states, only this mode is
\emph{uncompensated} by a counter-propagating state. For example, even
though they have different velocities, the left- and right-moving edge
states in the ($s$+$p$)-wave NCS still carry identical currents in
opposite directions at zero temperature. 
This statement can be formalized by examining~\eq{totcur}: Only the
hopping term is sensitive to the chiral structure, as it is
proportional to the difference between the number of states
(integrated LDOS) below the Fermi energy at $+k_y$ and
$-k_y$. Inspection of~\fig{fig2L} clearly shows that only the opposite
energy shift of the oppositely polarized nondegenerate flat bands can
generate such a number difference. The singular behavior of the
current immediately follows, as the number difference appears even for 
infinitesimal field strength, and does not change as the field is
increased. 
The hopping contribution is vanishing in all other cases, and  
the current is instead due to the SOC term in~\eq{totcur}, which is 
proportional to the sum of the $x$-spin polarization at $+k_y$ and
$-k_y$. This naturally connects the linear variation of the current with field
strength to the induced surface magnetization.

In the case of
the majority-singlet ($d_{xy}$+$p$)-wave NCS, however, the splitting of the
doubly degenerate flat bands for $|k_y|<k_{F,+}$
also gives a small singular response to the
exchange field, as even an infinitesimal shift of the negatively (positively)
spin-polarized states below (above) the Fermi energy generates a
finite $x$-spin polarization. The current
from the doubly degenerate states nevertheless increases monotonically with
exchange field strength, and so the overall decreasing current remains a
signature of the nondegenerate flat bands. The doubly degenerate flat bands
are also distinguished by their dependence on the exchange-field 
orientation. 
While the magnitude of the current  contributed by the doubly degenerate flat
bands  is equal for both $x$- and $z$-oriented  fields, the current due to the
nondegenerate flat bands 
undergoes a large change as the field is rotated form the $x$ to the
$z$ axis~\cite{SM}.

Although we have specified our discussion to the
case of an insulating ferromagnet, we note that an edge current is also
induced by placing the NCS in contact with a ferromagnetic
\emph{metal}~\cite{Brydon2013,Ren2013}. The observation of a large edge
current at the interface between an NCS and any
ferromagnet would therefore be
strong proof of the existence of nondegenerate flat bands.
Experimental detection of the edge currents 
should be possible in spite of the Meissner effect, which implies that
screening currents exactly compensate the 
edge currents in a large sample. 
However, whereas the edge-current density decays into the
bulk on the scale of the coherence
length $\xi_0$, the screening only 
builds up over the scale of the penetration depth $\lambda_{L}$. For
an extreme type-II superconductor, characteristic of
many NCSs~\cite{bauerSigristbook}, the screening currents will 
therefore be suppressed in a sample of width $W$ with $\xi_0 \ll W \ll
\lambda_{L}$.
This argument also holds for an engineered NCS~\cite{engineered}.

\emph{Summary}.---We have proposed a novel test of nondegenerate flat
bands at the edge of a topological NCS based on their response to
an exchange field. Specifically, we have shown that due to their strong spin
polarization, they acquire a chiral dispersion when placed in contact with a 
ferromagnet. The resulting current shows a characteristic singular dependence
upon the exchange field strength, and dominates the current due to  other
edge states, including doubly degenerate flat bands.

\emph{Acknowledgments}.---The authors thank M.~Sigrist for useful discussions.
A.P.S. thanks NORDITA for its hospitality and financial support.
C.T. acknowledges financial support by the Deutsche Forschungsgemeinschaft
through Research Training Group GRK 1621.

\balance

 \clearpage
\newpage

\appendix

\setcounter{figure}{0}
\makeatletter 
\renewcommand{\thefigure}{S\@arabic\c@figure} 

\makeatother

\begin{center}
\textbf{
\large{Supplemental Material for}}
\vspace{0.4cm} 

\textbf{
\large{
``Edge currents as a signature of flat bands in topological superconductors" } 
}
\end{center}

\vspace{0.1cm}

\begin{center}
\textbf{Authors:} Andreas P.\ Schnyder, Carsten Timm, and  P.~M.~R.~Brydon
\end{center}

\vspace{0.5cm}

For completeness, we present in this Supplement Material the 
$z$ component of the spin-LDOS at the edge of an NCS. We also
study the  changes in the edge band structure
induced by a $z$-polarized exchange field in the leading edge of the NCS
and determine the  resulting edge currents.
The results are qualitatively similar to the ones for
 an $x$-polarized exchange field, which
 we have discussed in the main text.

\section{I.~~~Edge band structure}

\begin{figure}[t!]
\includegraphics[clip,angle=0,width=\columnwidth]{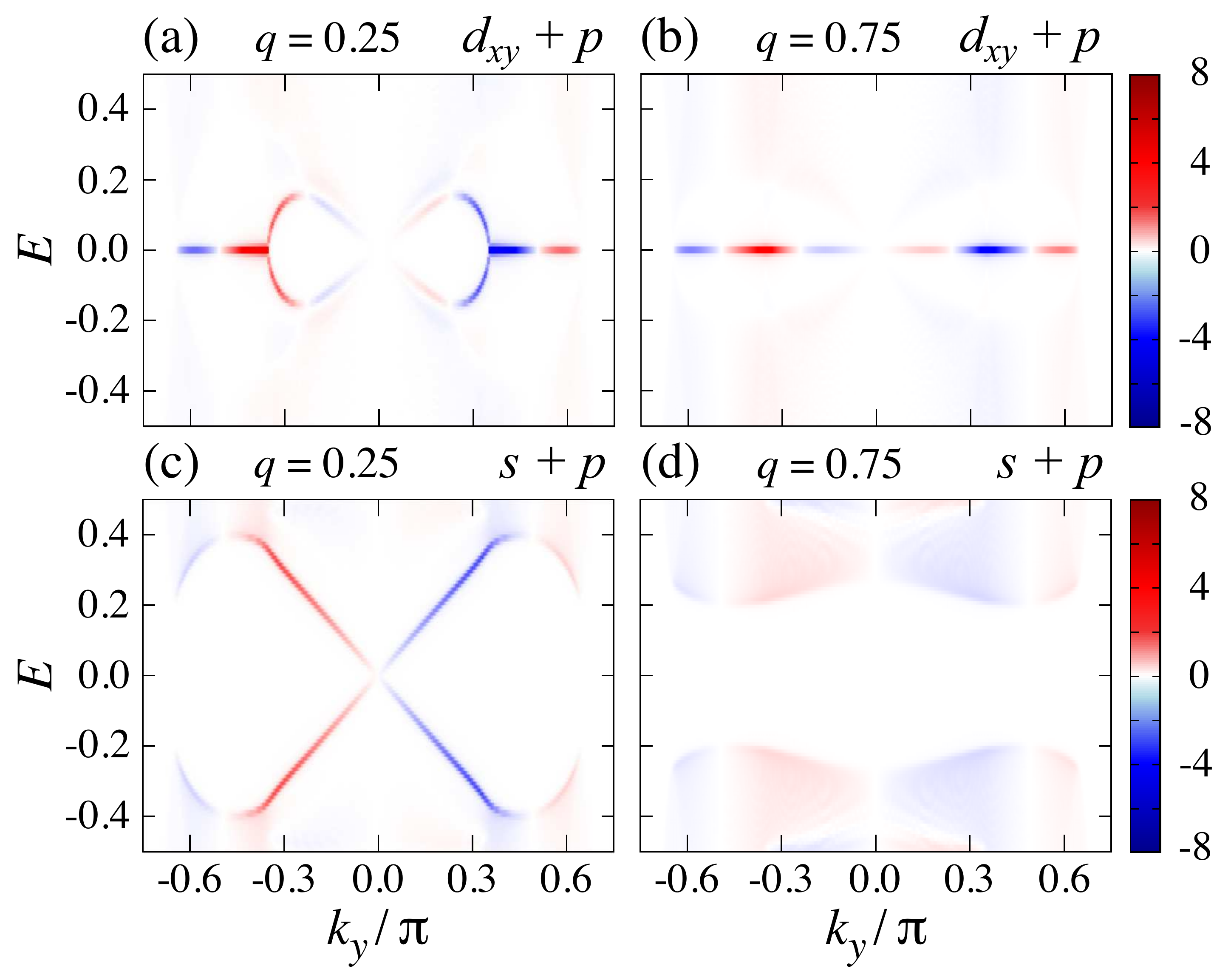}
\caption{\label{fig6L}
$z$ component of the energy- and momentum-resolved spin-LDOS for the outermost layer
of the ($d_{xy}$+$p$)-wave NCS with (a) $q=0.25$ and (b) $q=0.75$,
and for the ($s$+$p$)-wave NCS with (c) $q=0.25$ and (d) $q=0.75$.
As in Fig.~1 of the main text the spin-LDOS is plotted in units of $\hbar / 20$ on a linear scale.
}
\end{figure}

Fig.~\ref{fig6L}  displays the $z$
component of the energy- and momentum-resolved spin-LDOS for the outermost layer at the (10) edge
of the $(d_{xy}$+$p$)-wave and ($s$+$p$)-wave NCS.
This figure should be compared to
the $x$ component of the momentum-resolved spin-LDOS, which is
depicted in Figs.~1(b), (d), (f), and (h) of the main text. As stated in the main text, 
both the continuum and subgap states have a strong total spin polarization in the $xz$ spin-plane.
As required by
time-reversal symmetry, the total spin polarization 
is an odd function of edge momentum $k_y$.
The nondegenerate zero-energy flat bands of the ($d_{xy}$+$p$)-wave NCS show a
particularly strong and robust $x$-spin polarization [Figs.~1(b) and (d)], with a weaker $z$-spin
polarization that changes sign near $\pm(k_{F,-}-k_{F,+})/2$ [Figs.~\ref{fig6L}(a) and (b)].
The doubly degenerate states of the ($d_{xy}$+$p$)-wave NCS, which appear in the
region $|k_y| \in [k_{F,+}, k_{F,-}]$, give opposite contributions to the
spin-LDOS of unequal magnitude. They therefore exhibit a weaker
overall spin polarization than the nondegenerate states [Figs.~1(d) and \ref{fig6L}(b)].

\begin{figure}[t!]
\includegraphics[clip,angle=0,width=\columnwidth]{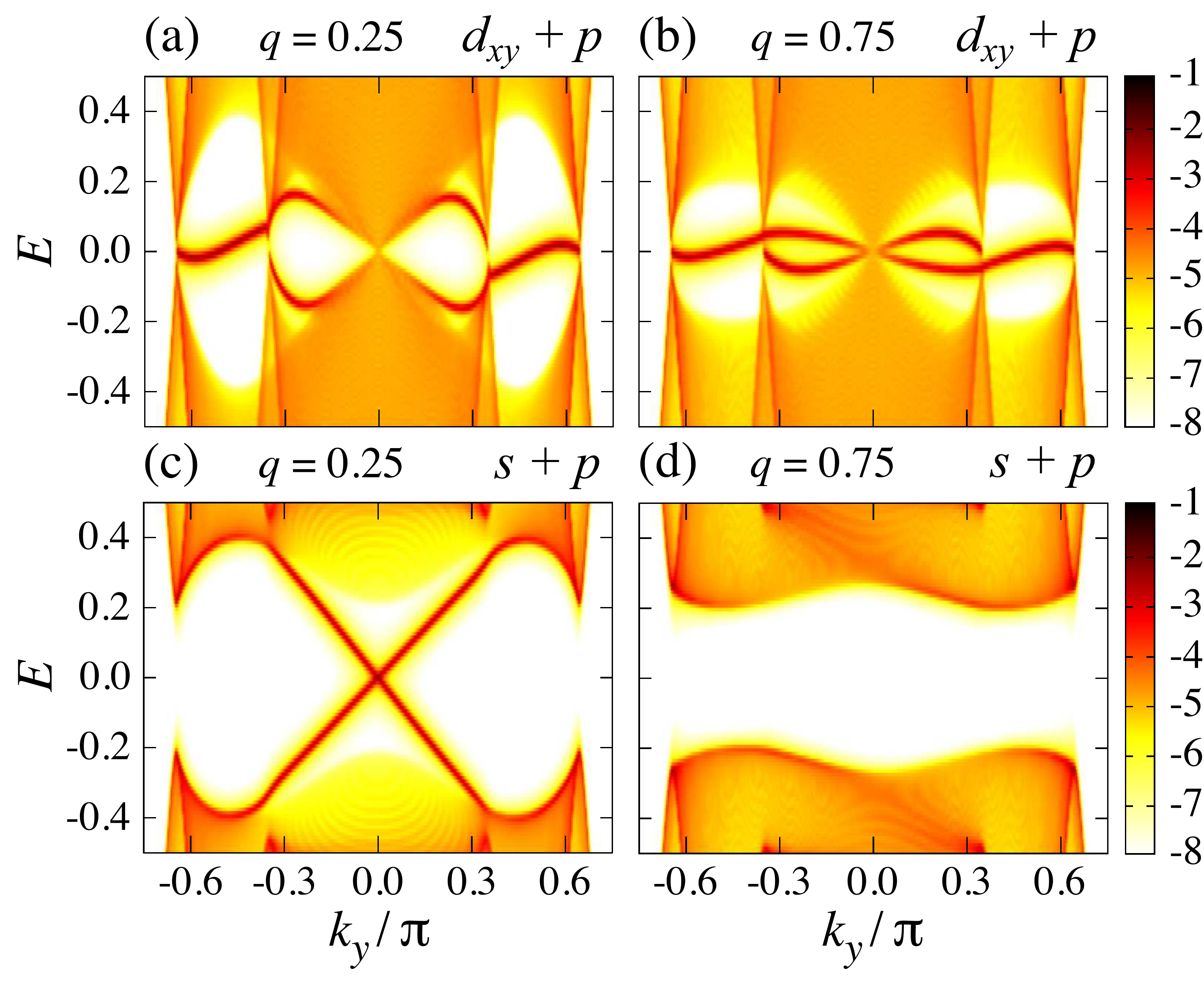}
\caption{\label{fig7L}
Energy- and momentum-resolved  LDOS on a 
log scale for the ten outermost layers at the (10) edge of (a), (b)
the ($d_{xy}$+$p$)-wave and (c), (d) the
($s$+$p$)-wave NCS 
in the presence of a $z$-polarized exchange field in the leading edge
with ${\bf H}_{\textrm{ex}}= 0.4 \, {\bf e}_z$.
In the left column we plot the results for $q=0.25$ (majority singlet), whereas
in the right column we have $q=0.75$ (majority triplet).
}
\end{figure}

An exchange field applied to the leading edge of a ($d_{xy}$+$p$)-wave NCS turns the nondegenerate flat bands
into dispersing chiral modes. We have demonstrated this for an $x$-polarized exchange field in Figs.~2(a) and (b) of the main text.
Figs.~\ref{fig7L}(a) and (b) show the corresponding results  for 
an exchange field polarized along the $z$ axis. As before, we find 
that  the exchange field ${\bf H}_{\textrm{ex}}= 0.4 \, {\bf e}_z$ leads to an energy shift
of the edge states which is proportional to  $  H^{z}_{\textrm{ex}}\,
\rho^{z}_{1}(E, k_y)$. Finally,  also the left- and right-moving edge
states of the ($s$+$p$)-wave NCS are shifted in energy due to the proximity-induced exchange field,
see Fig.~\ref{fig7L}(c) and compare with Fig.~2(c) in the main text.

%%%%%%%%%%%%%%%%%%%%%%%%%%%%
%%%%%%%%%%%%%%%%%%%%%%%%%%%%

%%%%%%%%%%%%%%%%%%%%%%%%%%%%
%%%%%%%%%%%%%%%%%%%%%%%%%%%%

\section{II.~~~Edge currents} 

\begin{figure}[t!]
\includegraphics[clip,angle=0,width=\columnwidth]{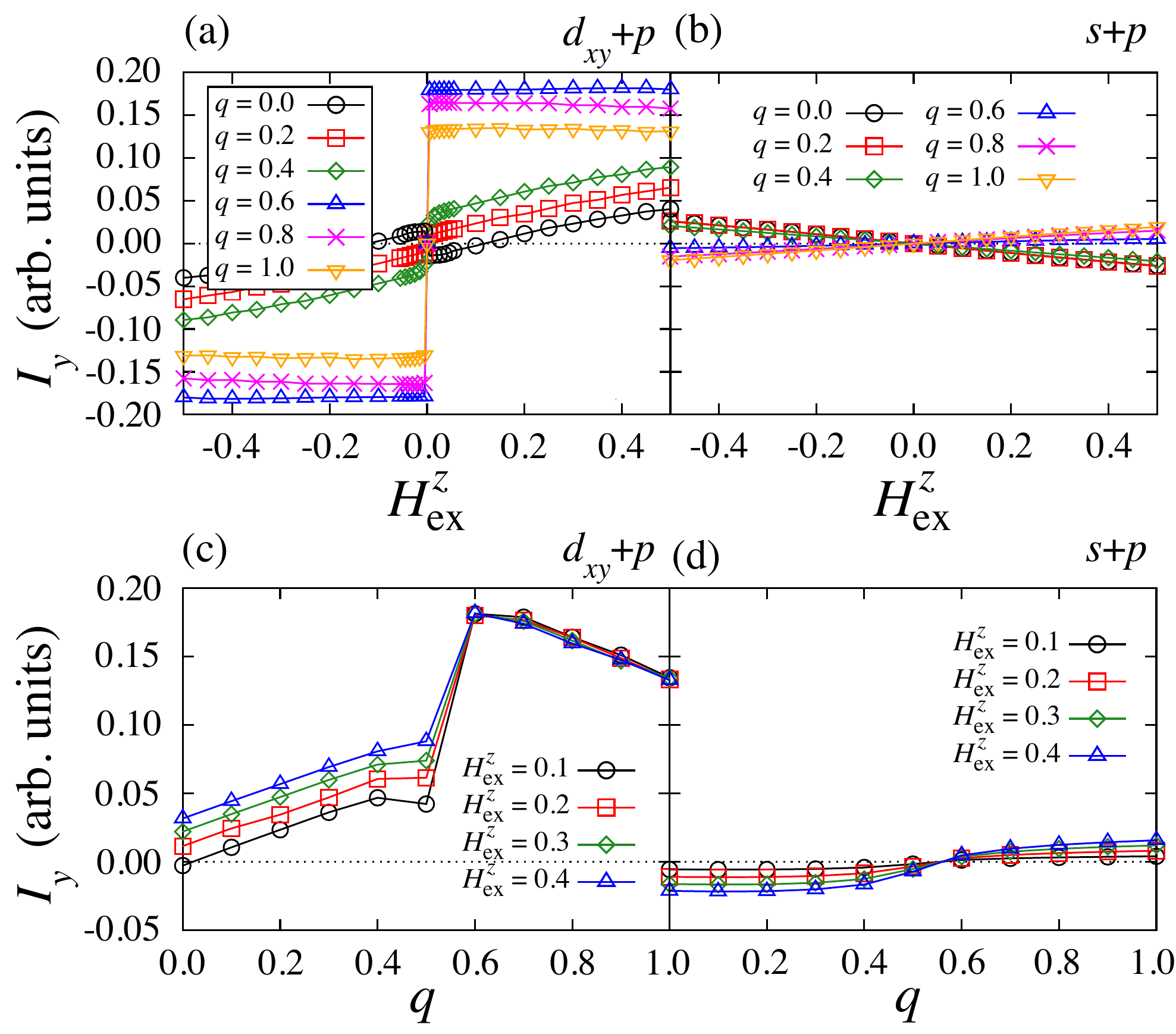}
\caption{\label{fig8L}
Top row:  Edge current $I_y$ as a function of exchange field
${\bf H}_{\textrm{ex}}=H^{z}_{\textrm{ex}} {\bf e}_z$ for various values of the singlet-triplet parameter $q$
for (a) the ($d_{xy}$+$p$)-wave and (b) the ($s$+$p$)-wave NCS.
Bottom row: Edge current $I_y$ as a function of  singlet-triplet parameter $q$
for various values of  the  exchange field ${\bf H}_{\textrm{ex}}=H^{z}_{\textrm{ex}} {\bf e}_z$ applied to the edge layer of
(c) the ($d_{xy}$+$p$)-wave and (d) the  ($s$+$p$)-wave NCS.}
\end{figure}

As explained in the main text, the energy shifts of the edge states induced by the exchange field
lead to a spontaneous  current at the edge of the NCS. 
We have exemplified this for an $x$-polarized exchange field in Fig.~3 of the main text.
In Fig.~\ref{fig8L} we present the edge current $I_y$  for a $z$-polarized exchange field, as a function of 
singlet-triplet parameter $q$ and
exchange field $ {\bf H}_{\textrm{ex}}  = H^z_{\textrm{ex}} {\bf e}_z$.
 Similar to Fig.~3, we 
find that also a $z$-polarized exchange field indcues a finite current  both for the
($d_{xy}$+$p$)-wave and the  ($s$+$p$)-wave NCS.
As in Fig.~3(a), the current at the edge of the ($d_{xy}$+$p$)-wave NCS
exhibits a singular dependence on exchange-field strength [Fig.~\ref{fig8L}(a)].
The current at the edge of the ($s$+$p$)-wave
NCS, on the other hand, grows linearly with the exchange field, see Fig.~\ref{fig8L}(b).

\begin{figure}[h!]
\includegraphics[clip,angle=0,width=\columnwidth]{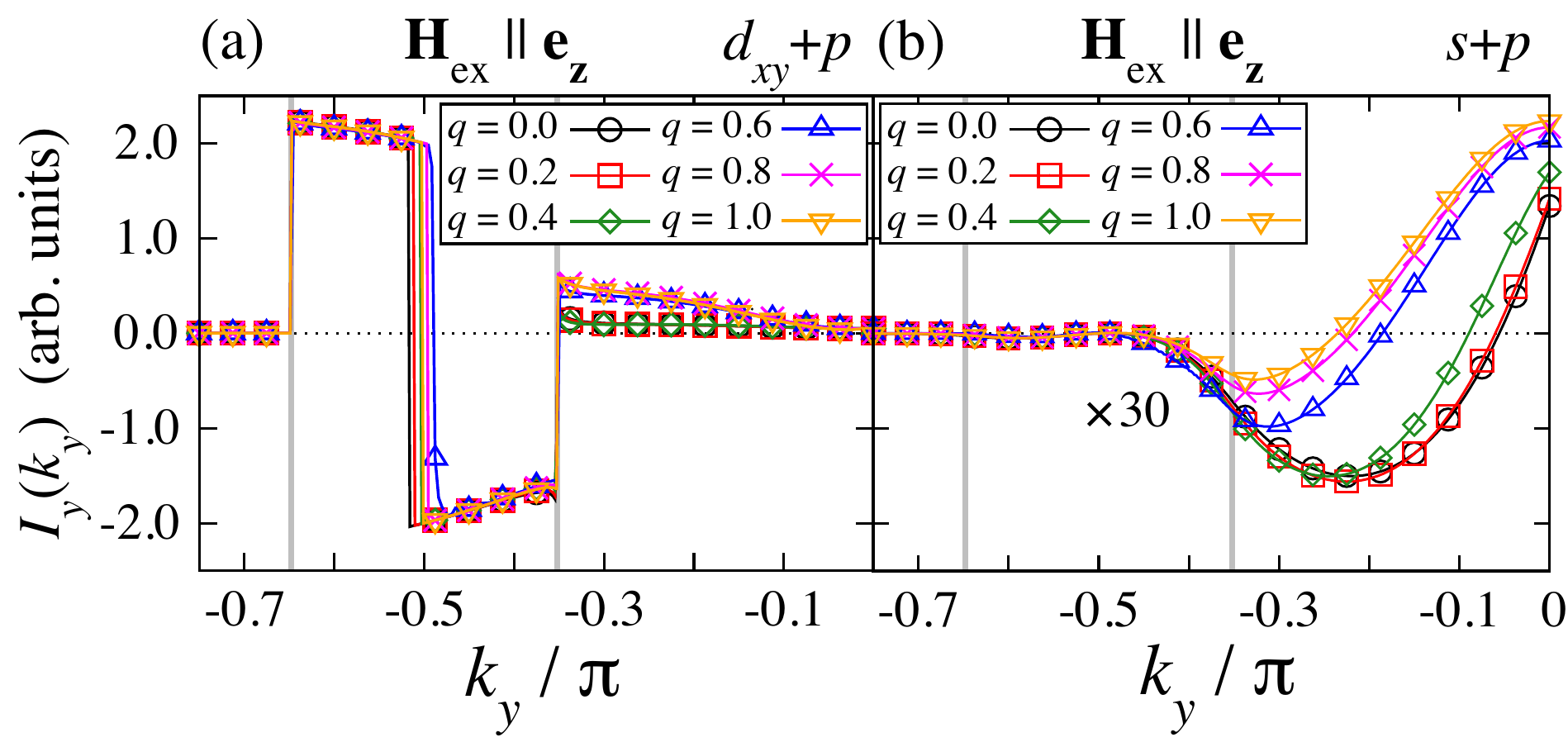}
\caption{\label{fig9L}
Momentum-resolved edge current $I_y ( k_y)$ for various values
of singlet-triplet parameter $q$ in (a) the ($d_{xy}$+$p$)-wave and
(b) the  ($s$+$p$)-wave NCS in the presence of an exchange field
${\bf H}_{\textrm{ex}}= 0.2 \, {\bf e}_z$ applied to the edge layer.
The current in the ($s$+$p$)-wave case has been multiplied 
by $30$ for better visibility.}
\end{figure}

Finally, in Fig.~\ref{fig9L} we present the momentum-resolved current $I_{y}(k_y)$
for an exchange field polarized along the $z$-axis. The corresponding plot
for the  $x$-polarized exchange field is given in Fig.~4 in the main text. 
Interestingly, for an exchange field along the $z$ axis the current
  shows a jump in $I_y(k_y)$ at the location of the sign
change of the $z$ component of the spin polarization of the nondegenerate
edge  states, cf.\ Figs.~\ref{fig6L}(a) and (b).

%%%%%%%%%%%%%%%%%%%%%%%%%%%%
%%%%%%%%%%%%%%%%%%%%%%%%%%%%

%%%%%%%%%%%%%%%%%%%%%%%%%%%%
%%%%%%%%%%%%%%%%%%%%%%%%%%%%

 \clearpage

\end{document}